\documentclass[sigconf,authorversion]{acmart}
\pdfoutput=1
\renewcommand\footnotetextcopyrightpermission[1]{} 
\usepackage{booktabs} 

\usepackage[T1]{fontenc}
\usepackage{lmodern}

\usepackage[absolute]{textpos}

\usepackage{ulem}
\usepackage{listings}
\usepackage{graphicx}
\usepackage{subcaption}

\newcommand{\asm}[1]{{\tt\normalsize#1}}
\newcommand{\code}[1]{{\tt\normalsize#1}}

\newcommand{\etal}[1]{#1 \emph{et al.}} 
\lstdefinestyle{customc}{
    breaklines=true,
    frame=L,
    numberstyle=\footnotesize\color{gray},
    numbersep=5pt,
    language=C,
    showstringspaces=false,
    basicstyle=\footnotesize\ttfamily,
    keywordstyle=\bfseries\color{green!40!black},
    commentstyle=\itshape\color{purple!40!black},
    identifierstyle=\color{blue},
    stringstyle=\color{orange},
}
\lstnewenvironment{ccode}[1][]{
    \lstset{
        style=customc,
        frame=single,
        #1
    }
}
{}

\lstdefinelanguage
[x64]{Assembler}     
[x86masm]{Assembler} 
{morekeywords={CDQE,CQO,CMPSQ,CMPXCHG16B,JRCXZ,LODSQ,MOVSXD, %
        POPFQ,PUSHFQ,SCASQ,STOSQ,IRETQ,RDTSCP,SWAPGS, %
        rax,rdx,rcx,rbx,rsi,rdi,rsp,rbp, %
        cpuid,rdtsc,%
        r8,r8d,r8w,r8b,r9,r9d,r9w,r9b, %
        r10,r10d,r10w,r10b,r11,r11d,r11w,r11b, %
        r12,r12d,r12w,r12b,r13,r13d,r13w,r13b, %
        r14,r14d,r14w,r14b,r15,r15d,r15w,r15b}} 

\lstdefinestyle{customasm}{
    keywords={cpuid, rdtsc, rdtscp},
    frame=L,
    xleftmargin=\parindent,
    language=[x64]Assembler,
    basicstyle=\small\ttfamily,
    keepspaces=true,
    commentstyle=\itshape\color{purple!40!black},
    numberstyle=\footnotesize\color{gray},
    numbersep=5pt
}
\lstnewenvironment{asmcode}[1][]{
    \lstset{
        style=customasm,
        basicstyle=\ttfamily\small,
        frame=single,
        #1
    }
}
{}

\lstdefinelanguage{JavaScript}{
keywords={break, case, catch, continue, debugger, default, delete, do, else, false, finally, for, function, if, in, instanceof, new, null, return, switch, this, throw, true, try, typeof, var, void, while, with},
    morecomment=[l]{//},
    morecomment=[s]{/*}{*/},
    morestring=[b]',
    morestring=[b]",
    ndkeywords={class, export, boolean, throw, implements, import, this},
    keywordstyle=\color{blue}\bfseries,
    ndkeywordstyle=\color{darkgray}\bfseries,
    identifierstyle=\color{black},
    commentstyle=\color{purple}\ttfamily,
    stringstyle=\color{red}\ttfamily,
    sensitive=true
}

\begin{document}

%

\fancyhead[C]{{\bf authors' version; not for redistribution}}
\fancyhead[L]{ret2spec}
\fancyhead[R]{\thepage}
\renewcommand{\headrulewidth}{0.4pt}

\title{~\\~\\ret2spec: Speculative Execution Using Return Stack Buffers}


\author{Giorgi Maisuradze}
\affiliation{%
    \institution{CISPA, Saarland University}
}
\email{giorgi.maisuradze@cispa.saarland}
\author{Christian Rossow}
\affiliation{%
    \institution{CISPA, Saarland University}
}
\email{rossow@cispa.saarland}

\renewcommand{\shortauthors}{G. Maisuradze and C. Rossow}

\begin{abstract}
Speculative execution is an optimization technique that has been part of CPUs for over a decade.
It predicts the outcome and target of branch instructions to avoid stalling the execution pipeline.
However, until recently, the security implications of speculative code execution have not been studied.

In this paper, we investigate a special type of branch predictor that is responsible for predicting return addresses.
To the best of our knowledge, we are the first to study return address predictors and their consequences for the security of modern software.
In our work, we show how return stack buffers (RSBs), the core unit of return address predictors, can be used to trigger misspeculations.
Based on this knowledge, we propose two new attack variants using RSBs that give attackers similar capabilities as the documented Spectre attacks.
We show how local attackers can gain arbitrary speculative code execution across processes, e.g., to leak passwords another user enters on a shared system.
Our evaluation showed that the recent Spectre countermeasures deployed in operating systems can also cover such RSB-based cross-process attacks.
Yet we then demonstrate that attackers can trigger misspeculation in JIT environments in order to leak arbitrary memory content of browser processes.
Reading outside the sandboxed memory region with JIT-compiled code is still possible with 80\% accuracy on average.
\end{abstract}

\begin{textblock}{13}(1.4,1)
    \noindent \LARGE \bf
        \textcopyright~Owner/Author 2018. This is the author's version of the work. It is posted here for your personal use. Not for redistribution. The definitive Version of Record was published in {Source Publication}, \url{http://dx.doi.org/10.1145/3243734.3243761}.
\end{textblock}


\maketitle

\section{Introduction}
\label{sec:introduction}

For decades, software has been able to abstract from the inner workings of operating systems and hardware, and significant research resources have been spent on assuring software security.
Yet only recently, the security community has started to investigate the security guarantees of the hardware underneath.
The first investigations were not reassuring, revealing multiple violations of security and privacy, e.g., demonstrating that cryptographic keys meant to be kept secret may leak via caching-based side channels~\cite{prime+probe,cache-missing-for-fun-and-profit,cache-timing-aes-bernstein}.
This recent discovery has piqued interest in the general topic of microarchitectural attacks.
More and more researchers aim to identify potential problems, assess their impact on security, and develop countermeasures to uphold previously-assumed security guarantees of the underlying hardware.
As a consequence, a variety of novel techniques have been proposed which abuse microarchitectural features, thereby making seemingly-secure programs vulnerable to different attacks~\cite{cache-template-attacks,spy-in-the-sandbox,cache-missing-for-fun-and-profit,cache-timing-aes-bernstein,rowhammerjs,rowhammer}.

One of the core drivers for recent microarchitectural attacks is the sheer complexity of modern CPUs.
The advancement of software puts a lot of pressure on hardware vendors to make their product as fast as possible using a variety of optimization strategies.
However, even simple CPU optimizations can severely threaten the security guarantees of software relying on the CPU.
Caching-based side channels are a notable example of this problem: such side channels exist since caches that improve the access time to main memory are shared across processes.
Thus, caching can result in leaking cryptographic keys~\cite{cache-missing-for-fun-and-profit,cache-timing-aes-bernstein}, key-stroke snooping, or even eavesdropping on messages from secure communications~\cite{cache-template-attacks,spy-in-the-sandbox}.

Besides caching, modern CPUs deploy several other optimization techniques to speed up executions, two of which we will study in more detail.
First, in out-of-order execution, instead of enforcing a strict execution order of programs, CPUs can reorder instructions, i.e., execute new instructions before older ones if there are no dependencies between them.
Second, in speculative execution, CPUs predict the outcome/target of branch instructions.
Both these strategies increase the utilization of execution units and greatly improve the performance.
However, they also execute instructions in advance, meaning they can cause instructions to execute that would have not been executed in a sequential execution sequence.
For example, it can happen that an older instruction raises an exception, or that the branch predictor mispredicts.
In this case, the out-of-order executed instructions are rolled back, restoring the architectural state at the moment of the fault (or misspeculation).
Ideally, the architectural state is the same as in strict sequential execution.
However, this is not the case: instructions executed out of order can influence the state in a manner that can be detected.
Meltdown~\cite{meltdown} and Spectre~\cite{spectre} are great examples of this class of problems.
Meltdown exploits a bug in Intel's out-of-order engine, allowing the privileged kernel-space data to be read from unprivileged processes.
Spectre poisons the branch target buffer (BTB) and thus tricks the branch prediction unit into bypassing bounds checks in sandboxed memory accesses, or even triggering arbitrary speculative code execution in different processes on the same core.
To mitigate these threats, operating systems had to make major changes in their design (e.g., isolating the kernel address space from user space~\cite{kaiser}), and hardware vendors introduced microcode updates to add new instructions to control the degree of the aforementioned CPU optimization techniques~\cite{intel-spectre-whitepaper}.

In this paper, we further investigate speculative execution and show that attacks are possible beyond the already-documented abuse of BTBs.
More specifically, we look into the part of branch prediction units that are responsible for predicting return addresses.
Since they are the core of the return address predictor, we will in particular investigate the properties of return stack buffers (RSBs).
RSBs are small microarchitectural buffers that remember return addresses of the most recent calls and speed up function returns.
Given that return addresses are stored on the stack, without such RSBs, a memory access is required to fetch a return destination, possibly taking hundreds of cycles if retrieved from main memory.
In contrast, with RSBs, the top RSB entry can be read instantaneously.
RSBs thus eliminate the waiting time in the case of a correct prediction, or in the worst case (i.e., in case of a misprediction) face \textit{almost}\footnote{Rolling back the pipeline on misspeculation adds an overhead of a few cycles.} no additional penalty.

Despite being mentioned as \emph{potential} threat in the initial report from Google Project Zero~\cite{meltdown-p0} and Spectre~\cite{spectre}, the security implications of abusing RSBs have not yet been publicly documented, and only very recent studies have started to investigate timing implication of return address mispredictions at all~\cite{stuffedcow-rsb-microbenchmarks}.
To the best of our knowledge, we are the first to systematically study and demonstrate the \emph{actual} security implications of RSBs.
We furthermore show the degree to which attackers can provoke RSB-based speculative execution by overflowing the RSB, by crafting malicious RSB entries prior to context switches, or by asymmetric function call/return pairs.

Based on these principles, we provide two RSB-based attack techniques that both allow attackers to read user-level memory that they should not be able to read.
In the first attack (Section~\ref{sec:attack1}), we assume a local attacker that can spawn arbitrary new programs that aim to read another user's process memory.
To this end, we show how one can poison RSBs to force the colocated processes (on the same logical core) to execute arbitrary code speculatively, and thus report back potential secrets.
Interestingly, operating systems (coincidentally) mitigate this attack by flushing RSBs upon context switches.
Yet these mitigations were introduced to anticipate potential RSB underflows that trigger the BTB\footnote{https://patchwork.kernel.org/patch/10150765/}, possibly leading to speculatively executing user-mode code with kernel privileges.
RSB flushing is thus only used when either SMEP\footnote{SMEP (Supervisor Mode Execution Protection) is a recent x86 feature implemented by Intel to protect against executing code from user-space memory in kernel mode.} is disabled or when CPUs of generation Skylake+ are used.
This hints at the fact that RSB stuffing was introduced in order to avoid speculative execution of user-land code from kernel (non-SMEP case) or BTB injections (Skylake+ CPUs fall back to BTB predictions on RSB underflow).
However, as this defense is not always active, several CPU generations are still vulnerable to the attack demonstrated in this paper.
Our work shows that RSB-based speculated execution (i) can indeed be provoked by local attackers with non-negligible probability, and (ii) goes beyond the currently-assumed problem of falling back to the BTB (thus allowing for Spectre) when underflowing RSBs, and thus, can be generalized to the non-trustworthiness of attacker-controlled RSBs.

In our second attack (Section~\ref{sec:attack2}), we investigate how attackers can abuse RSBs to trigger speculation of arbitrary code inside the same process---notably \emph{without} requiring a context switch, and thus effectively evading the aforementioned defense.
We assume an attacker that controls a web site the target user visits, and by carefully crafting this web site, aims to read memory of the victim's browser process.
Technically, we leverage just-in-time (JIT) compilation of WebAssembly to create code patterns that are not protected by memory isolation techniques and thus can read arbitrary memory of a browser process.
By doing so, we show that adversaries can bypass memory sandboxing and read data from arbitrary memory addresses.

Both attack types demonstrate that speculative execution is not limited to attackers penetrating the BTB.
While our attacks result in similar capabilities as Spectre, the underlying attack principles to manipulate the RSB are orthogonal to the known poisoning strategies.
We thus also discuss how existing and new countermeasures against RSB-based attacks can mitigate this new risk (Section~\ref{sec:discussion}).
We conclude the paper with vendor and developer reactions that we received after responsibly disclosing the internals of this new threat.

In this paper, we provide the following contributions:
\begin{itemize}
	\item We study the return address predictor, an important yet so far overlooked module in the prediction unit.
	To the best of our knowledge, we are the first to demonstrate the actual abuse potential of RSBs.
	
	\item We propose attack techniques to trigger misspeculations via the RSB.
	This can be useful in future studies that will target speculative code execution.
	In contrast to using the branch predictor, which requires a prior training phase, RSBs can be forced to misspeculate to required addresses without prior training.
	
	\item We then propose cross-process speculative execution of arbitrary code (similar to Spectre/Variant 1).
	We evaluate the results by leaking keystrokes from a specially-crafted \texttt{bash}-like program.
	Using our synthetic program example, we demonstrate that such attacks are in principle conceivable, showing the importance of applying the existing OS-based defenses to every microarchitecture in order to fully mitigate our attack.
	
	\item Finally, we show how to trigger misspeculations via RSBs in JIT-compiled code.
	We leverage this to execute arbitrary code speculatively and, by doing so, bypass memory sandboxing techniques, allowing arbitrary memory reads.
	We evaluate our technique in Firefox 59 (with a modified timer for higher precision).
\end{itemize}

\section{Background}
\label{sec:background}

In the following, we will present the key features of x86 that are important to understand for the remainder of this paper.
While similar concepts are also common in other architectures, for brevity and due to its popularity, we focus on x86.

\subsection{Out-of-Order Execution}
\label{sec:background-ooo-execution}
Being a CISC (Complex Instruction Set Computing) architecture, x86 has to support a multitude of instructions.
Implementing all such instructions in circuits would require an enormous amount of transistors, thus also drastically increasing the power consumption.
Therefore, under the hood, both main manufacturers of x86 CPUs (Intel and AMD) use micro-OPs, which can be seen as a simplified RISC (Reduced Instruction Set Computing) machine that runs inside the CPU.
All instructions from the x86 ISA are then dynamically decoded into their corresponding micro-OPs, and are then executed on much simpler execution units.
This allows manufacturers to reduce the number of required execution paths, decreasing both production cost and power consumption.

Having a variety of different instructions, sequential execution becomes another bottleneck.
The concept of splitting up complex instructions into smaller operations also makes it possible to reorder the execution of micro-OPs to gain performance.
In a strict sequential execution, an instruction $N$ cannot be started unless all preceding instructions, $1..N-1$, are finished executing.
This is especially problematic for instructions with completely different timing properties, e.g., zeroing a register and reading a value from main memory.
Out-of-order execution deals with this issue by executing instructions out of order, provided they do not depend on one another.

To implement out-of-order execution, x86 maintains a so-called reorder buffer (ROB), which keeps a FIFO buffer of micro-OPs in their original order, while executing them out of order.
If a micro-OP is in the ROB it (i) is waiting for its dependencies to be resolved, (ii) is ready to be executed, (iii) is already being executed, or (iv) is done executing but was not yet committed.
Committing (also called \emph{retiring}) a micro-OP reflects its changes back to the architectural state, e.g., modifying the architectural (ISA-visible) registers or writing data back to memory.
Given that the programs assume a strict sequential order, the ROB commits instructions in order such that the architectural state is updated sequentially.

\subsection{Speculative Execution}
\label{sec:background-speculative-execution}

Modern CPUs augment out-of-order execution with an orthogonal feature called speculative execution.
The key observation here is that while executing instructions, CPUs can encounter a branch instruction that depends on the result of a preceding instruction.
This would never happen in a strict sequential (non-parallel) execution, as all previous instructions before the branch would have been resolved.
To cope with this problem in modern out-of-order CPUs that execute multiple instructions in parallel, the simplest solution is to wait until the branch condition/target is resolved, and only then continue the execution.
However, this would serialize execution at branch instructions, which would degrade the performance, especially given the high number of branch instructions in x86.

Speculative execution represents an efficient alternative solution and is consequently used in all modern CPUs.
Speculative execution uses a branch prediction unit (BPU), which predicts the outcome of conditional branch instructions (i.e., taken/not taken).
The out-of-order engine then continues execution on the predicted path.
This general concept is not limited to direct branches that always have a fixed jump target.
For example, consider indirect branches (such as indirect calls and jumps) that need to be resolved before being executed, i.e., the branch target can be stored either in a register or in memory.
In this case, the branch destination is the value that needs to be predicted.
To support indirect branches, the branch target buffer (BTB) stores a mapping between the branch source and its likely destination.

The two recently disclosed microarchitectural attacks, Spectre and Meltdown, abuse the aforementioned out-of-order and speculative execution engines.
Meltdown uses the fact that out-of-order engines do not handle exceptions until the retirement stage, and leverages it to access memory regions that would otherwise trigger a fault (e.g., kernel memory).
In Meltdown, authors exploit the bug in Intel's out-of-order execution engine, which reveals the data from the faulty memory access for a few cycles.
This time, however, is enough to do a dependent memory access on the data.
Although the dependent memory access will be flushed from the pipeline after handling the fault, the cache line for that address will remain cached, thus creating a side channel for leaking the data.
Conversely, Spectre uses legitimate features of branch predictors to mount an attack:
it mistrains the BPU for conditional branches in Variant 1, and injects BTB entries with arbitrary branch targets in Variant 2.
Variant 1 can be used to bypass bounds checking and thus read outside the permitted bounds, while Variant 2 allows cross-process BTB injection, allowing arbitrary speculative execution of the code in other processes on the same physical core.

\subsection{Return Stack Buffers}
\label{sec:background-RSB}

A return instruction is a specific type of indirect branch that always jumps to the top element on the stack (i.e., translated to \asm{pop tmp; jmp tmp}).
Consequently, in principle, BTBs can also be used here as a generic prediction mechanism.
However, given that functions are called from multiple different places, BTBs will frequently mispredict the jump destination.
To increase the prediction rate, hardware manufacturers rely on the fact that call and return instructions are always executed in pairs.
Therefore, to predict the return address at a return site, CPUs remember the address of the instruction following the corresponding call instruction.
This prediction is done via return stack buffers (RSBs) that store the $N$ most recent return addresses (i.e., the addresses of instructions after the $N$ most recent calls).
Note that, in case of hyperthreading, RSBs are dedicated to a logical core.
The RSB size, $N$, varies per microarchitecture.
Most commonly, RSBs are $N=16$ entries large, and the longest reported RSB contains $N=32$ entries in AMD's Bulldozer architecture~\cite{agner-microarchitecture}.
In this paper, we assume an RSB size of 16, unless explicitly stated otherwise, but our principles also work for smaller or larger RSBs.

RSBs are modified when the CPU executes a call or return instruction.
Calls are simple: a new entry (the address of the next instruction) is added to the RSB and the top pointer is incremented.
If the RSB was already full, the oldest entry will be discarded.
Conversely, in case the of a return instruction, the value is taken from the RSB, the top pointer is decremented, and the read value is used for prediction.

Due to the their limited size, it naturally happens that the RSBs cannot fit all the return addresses.
For example, $N+1$ calls followed by $N+1$ returns will underflow the RSB in the last return instruction.
The way such an underflow is handled depends on the microarchitecture.
There are the following possible scenarios: (a) stop predicting whenever the RSB is empty, (b) stop using the RSB and switch to the BTB for predictions, and (c) use the RSB as a ring buffer and continue predicting (using $idx\%N$ as the RSB top pointer).
Out of these scenarios, (a) is the easiest to implement; however, it stops predicting return addresses as soon as the RSB is empty.
The prediction rate is improved in (b), which incorporates the BTB to predict return destinations.
However, the improvement is made at the expense of BTB entries, which might detriment other branches.
Finally, (c) is an optimization for deep recursive calls, where all RSB entries return to the same function.
Therefore, no matter how deep the recursion is, returns to the recursive function will be correctly predicted.
According to a recent study~\cite{stuffedcow-rsb-microbenchmarks}, most Intel CPUs use the cyclic behavior described in variant (c), while AMD's use variant (a) and stop prediction upon RSB underflows.
Intel microarchitectures after Skylake are known to use variant (b)\footnote{https://patchwork.kernel.org/patch/10150765/}.
Throughout the paper, we will refer to (c) as cyclic, and (a) and (b) as non-cyclic.

\section{General Attack Overview}
\label{sec:attack-general}

\begin{figure*}[ht]
	\centering
	\begin{subfigure}{.48\linewidth}
		\centering
		\includegraphics[bb=0 0 600 600,trim=0 160 110 0,clip,width=1.0\columnwidth]{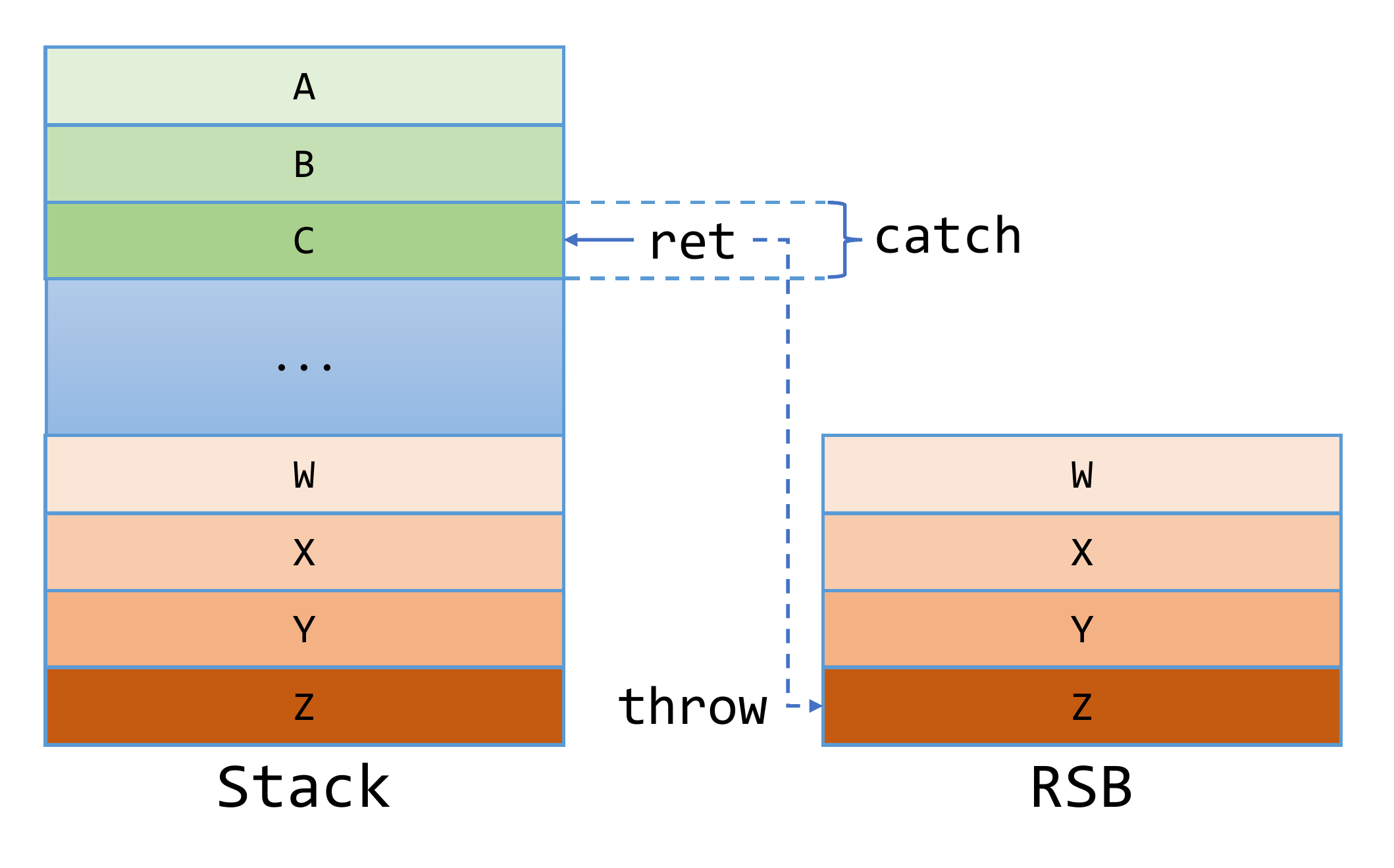}
		\captionsetup{width=.97\columnwidth}
		\caption{Exception handling: While the RSB predicts a return to function \texttt{Z}, the exception is caught by function \texttt{C}, causing a chain of misspeculations when \texttt{C} returns, as the RSB is misaligned to the return addresses on the stack.}
		\label{fig:rsb-misspec-exception}
	\end{subfigure}
	\begin{subfigure}{.48\linewidth}
		\centering
		\includegraphics[bb=0 0 600 600,trim=0 160 110 0,clip,width=1.0\columnwidth]{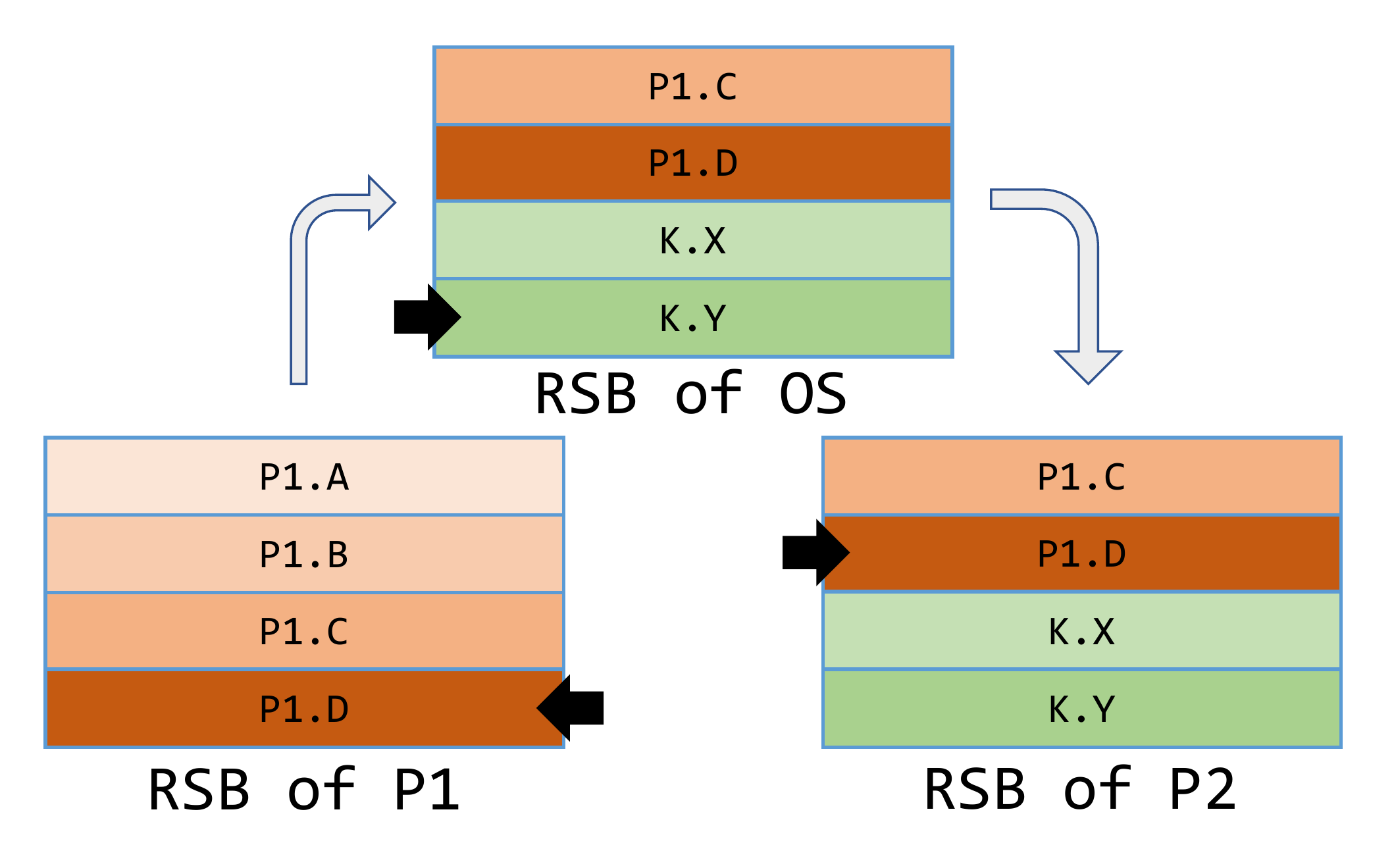}
		\captionsetup{width=.97\columnwidth}
		\caption{Context switch: When the kernel switches from process P1 to P2, the kernel will only evict a few entries with kernel-internal functions. After the context switch, P2 may thus mispredict and return to the remaining RSB entries that were added by P1.}
		\label{fig:rsb-misspec-contextswitch}
	\end{subfigure}\\
	\begin{subfigure}{.48\linewidth}
		\centering
		\includegraphics[bb=0 0 600 600,trim=0 160 110 0,clip,width=1.0\columnwidth]{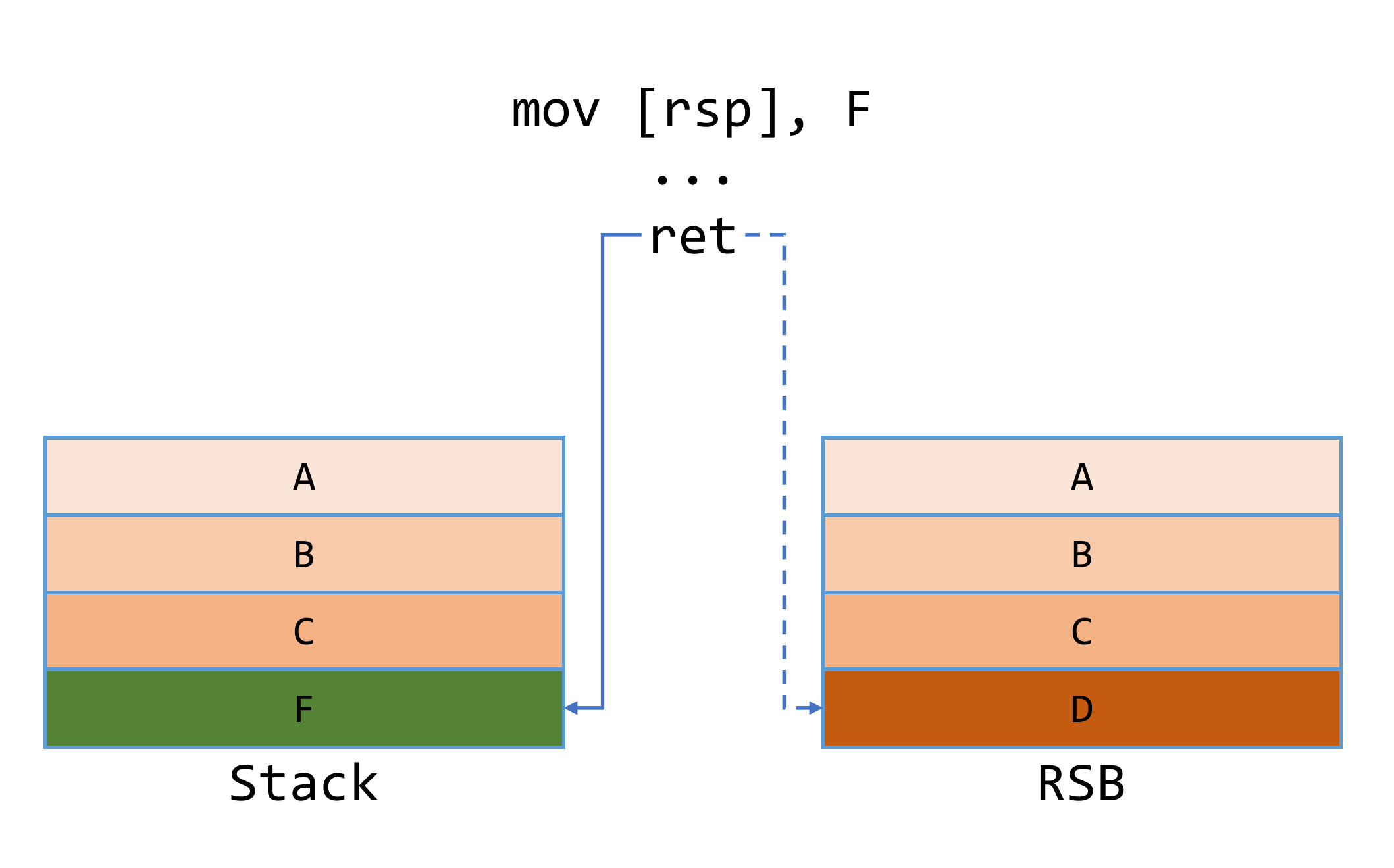}
		\captionsetup{width=.99\columnwidth}
		\caption{Direct overwrite: A process can enforce return mispredictions by replacing return addresses stored on the stack.}
		\label{fig:rsb-misspec-overwrite}
	\end{subfigure}
	\begin{subfigure}{.48\linewidth}
		\centering
		\includegraphics[bb=0 0 600 600,trim=0 160 110 0,clip,width=1.0\columnwidth]{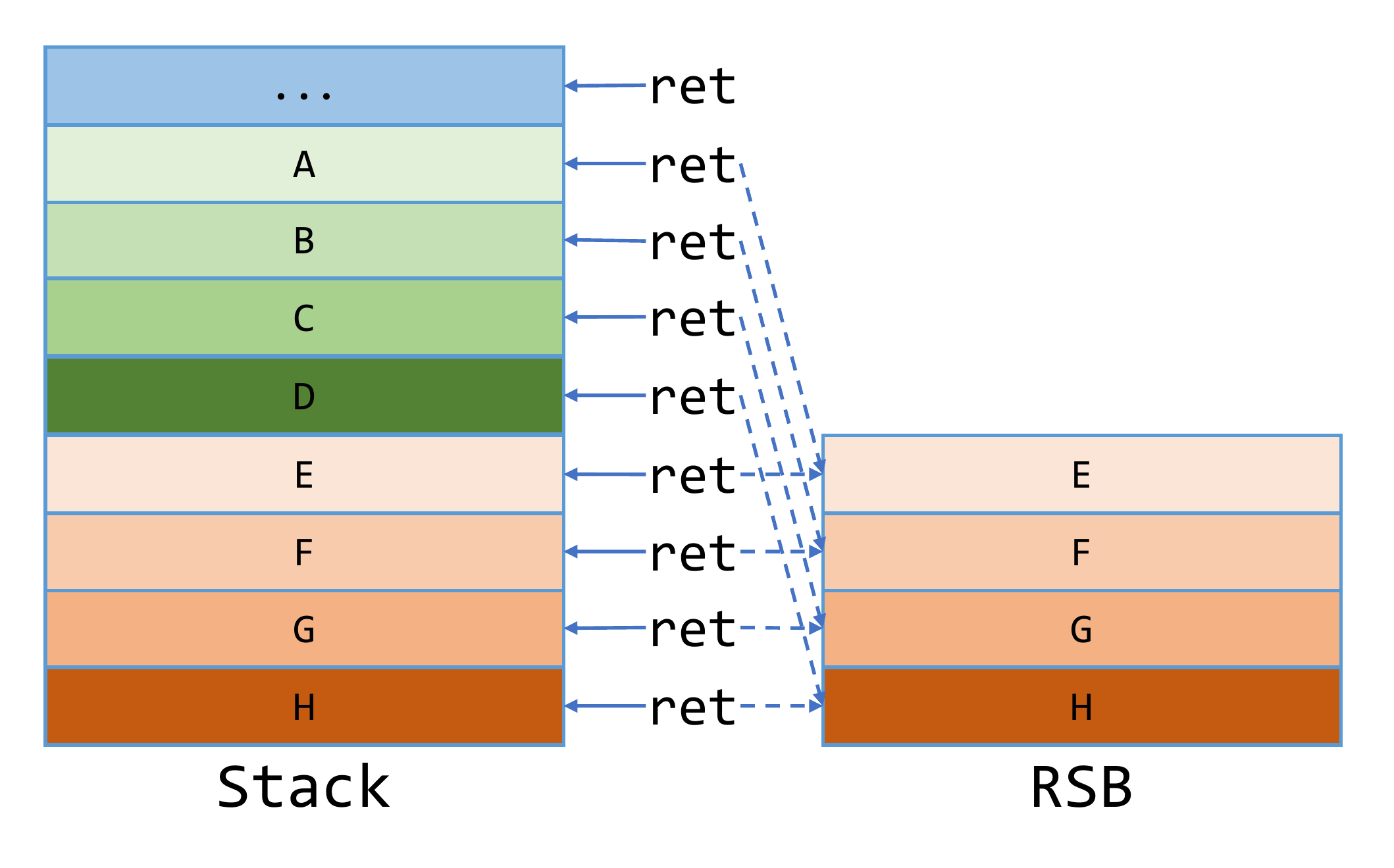}
		\captionsetup{width=.99\columnwidth}
		\caption{Circular RSB: After returning $N=4$ times, the predictor cycles over and will repeat the same prediction sequence of return addresses.}
		\label{fig:rsb-misspec-circular}
	\end{subfigure}
	\caption{Ways to enforce RSB misspeculation. We reduced the RSB size to $N=4$ entries for readability. The bold arrow points to the top element of each RSB. Thin solid arrows indicate actual returns, thin dashed arrows speculated returns.}
	\label{fig:rsb-misspeculations}
\end{figure*}

Before detailing specific attack scenarios, in this section, we introduce the basics of how RSB-based speculative execution can be achieved and be abused.
We explore whether and how attackers may manipulate the RSB entries in order to leak sensitive data using speculative execution that they could not access otherwise.
Similar to recent microarchitectural attacks~\cite{meltdown,spectre,speculose,sgxpectre,branchscope}, we trick the CPU to execute instructions that would not have been executed in a sequential execution.
The goal is to leak sensitive information in speculation, e.g., by caching a certain memory area that can be detected in a normal (non-speculative) execution.
The general idea of our attack can be divided into three steps:

\begin{description}
	\item[(A1)] trigger misspeculations in the return address predictor, i.e., enforce that returns mispredict
	\item[(A2)] divert the speculative execution to a known/controlled code sequence with the required context
	\item[(A3)] modify the architectural state in speculation, such that it can be detected from outside
\end{description}

\textbf{(A1) Triggering Misspeculation:}
From an attacker's perspective, enforcing that the return predictor misspeculates upon function return is essential to reliably divert speculative execution to attacker-controlled code (see A2 for how to control the speculated code).
Misspeculations can be achieved in several ways, depending on the RSB's underflow behavior~(as discussed in Section~\ref{sec:background-RSB}).

\emph{Non-Cyclic RSB:}
If the RSB stops speculating upon underflow, triggering a misspeculation will require abnormal control flow that violates the common assumption that functions return to their caller.
Some examples of such abnormalities are:
(i) exception handling, i.e., a try-catch block in the upper call stack and throwing an exception in a lower one (Figure~\ref{fig:rsb-misspec-exception});
(ii) a \code{setjmp/longjmp} pair, i.e., saving the current execution context at the time of calling \code{setjmp}, and restoring it at any later point when (\code{longjmp}) is called (the stack layout will be similar to Figure~\ref{fig:rsb-misspec-exception}, only with \code{setjmp/longjmp} instead of \code{try-catch/throw});
(iii) a context switch from one process to another, where the process being evicted was doing a chain of calls, while the scheduled process will do a sequence of returns (Figure~\ref{fig:rsb-misspec-contextswitch}); and
(iv) a process that deliberately overwrites the return address to the desired destination and then returns (Figure~\ref{fig:rsb-misspec-overwrite}).
Unsurprisingly, (iv) is not commonly used; however, it can be helpful for testing RSBs and triggering the misspeculation on demand.
In fact, in contrast to branch predictors, which require training to trigger misspeculation, RSBs can be forced to misspeculate with just a single store instruction (\asm{mov [rsp], ADDRESS; ret}, as in Figure~\ref{fig:rsb-misspec-overwrite}).

\emph{Cyclic RSB:} 
If RSBs are cyclic, misspeculation can---in addition to the methods mentioned before---be triggered by overflowing the RSB.
Figure\ref{fig:rsb-misspec-circular} depicts a scenario in which function \code{A} calls \code{B}, function \code{B} calls \code{C}, and so on.
Being limited in size ($N=4$ in this example), the RSB only contains the 4 most recently added return addresses.
Therefore, after correctly predicting four returns, when returning from \code{E}, the RSB will mispredict \code{H} as the return address instead of \code{D}.

Cyclic RSBs can also be leveraged and prepared by recursive functions.
For example, if we have two recursive functions \code{A} and \code{B}, and we call them in the following order:
\begin{itemize}
	\item \code{A} calls itself recursively $N_A$ times,
	\item in its deepest recursion level, \code{A} then calls \code{B}
	\item \code{B} calls itself recursively 16 times (size of the RSB)
\end{itemize}
then the first 16 returns, from \code{B}, will be predicted correctly.
However, the remaining $N_A$ returns will be mispredicted, and \code{B}'s call site will be speculated instead of \code{A}'s.

\textbf{(A2) Diverting Speculative Execution:}
Being able to trigger a misspeculation, the next step is to control the code that is executed speculatively.
Generally, misspeculation means that instructions from one function (e.g., \code{B}) are speculatively executed within the context of another (e.g., \code{A}).
As a simple example, consider a function that returns a secret value in \asm{rax}.
If this function is predicted to return to code that accesses attacker-accessible memory relative to \asm{rax}, this will leak the secret value.
Ideally, we control both, the context and the speculated code; however, having either one or the other can also be sufficient for a successful exploitation.

Let function \code{B} return and trigger a misspeculation in \code{A} (right after the call to \code{B}).
In the ideal case, we control the code that is misspeculated in \code{A}, and the context (i.e., the contents of the registers) in \code{B}.
Combining them together allows us to execute arbitrary code speculatively.
This will be the case for our attack in Section~\ref{sec:attack2}.
Another, more complicated case is when the context is fixed, e.g., the values of some registers are known, and we are also limited with the possibly-speculated code, e.g., it can be chosen from existing code pieces.
In this case, the challenge is to find code gadgets that use the correct registers from the context to leak their values.
For example, if we know that \asm{r8} contains a secret, we need to find a gadget that leaks \asm{r8}.
This case will be shown in Section~\ref{sec:attack1}.

\textbf{(A3) Feedback Channel:}
Finally, being able to execute arbitrary code in speculation, we have to report back the results from within the speculative execution to the normal execution environment.
To this end, similar to several side channels proposed in the past~\cite{prime+probe,flush+reload,cache-template-attacks}, we use secret-dependent memory accesses that modify the caching state.
Technically, if we want to leak the value in \asm{rax}, we read attacker-accessible memory using the secret value as offset, e.g., \asm{shl rax,12; mov r8,[rbx+rax]}.
This will result in caching the corresponding memory address (\asm{rbx+rax*4096}, where 4096 bytes is the page size).
Therefore, identifying the index of the cached page from \asm{rbx} will reveal the value of \asm{rax}.

The adversary can then use existing side channel techniques to observe these cache changes, such as Flush+Reload~\cite{flush+reload} or Prime+Probe~\cite{prime+probe}.
Flush+Reload is most accurate, but requires that the attacker and victim processes share memory.
Typically this is granted, given that operating systems share dynamic libraries (e.g., \code{libc}) to optimize memory usage.
Alternatively, Prime+Probe~\cite{prime+probe} works even without shared memory.
Here, the attacker measures whether the victim evicts one of the attacker-prepared cache lines.
By detecting the evicted cache line, the attacker can leak the bits corresponding to the cache line address.

\section{Cross-Process Speculative Exec.}
\label{sec:attack1}

In this section, we will describe how an attacker can abuse the general attack methodology described in the previous section to leak sensitive data from another process.
In Section~\ref{sec:attack2}, we will describe RSB-based attacks in scripting environments to read memory beyond the memory bounds of sandboxes.

\subsection{Threat Model}
\label{sec:attack1-threatmodel}

In the following, we envision a local attacker that can execute arbitrary user-level code on the victim's system.
The goal of the attacker is to leak sensitive data from another process (presumably of a different user) on the system, e.g., leaking input fed to the target process.
In our concrete example, we target a command line program that waits for user input (character-by-character), i.e., a blocking \code{stdin}, and we aim to read the user-entered data.
This setting is in line with Linux programs such as \code{bash} or \code{sudo}.
The attack principle, however, generalizes to any setting where attackers aim to read confidential in-memory data from other processes (e.g., key material, password lists, database contents, etc.).

For demonstration purposes, we assume that the kernel does not flush RSBs upon a context switch.
For example, this can be either an older kernel before without such patches, or an up-to-date kernel using an unprotected microarchitecture.
Furthermore, we assume that the victim process contains all attacker-required gadgets.
In our example, we simply add these code pieces to the victim process.
Finally, we assume that ASLR is either disabled or has been broken by the attacker.

\subsection{Triggering Speculative Code Execution}
\label{sec:attack1-rsb-cross-process-details}

We now apply the general attack principles to the scenario where an adversarial process executes alongside a victim process.
The attacker aims to trigger return address misprediction in the victim's process, and divert the speculative control flow to an attacker-desired location.
The fact that victim and attacker are in different processes complicates matters, as the context of the execution (i.e., the register contents) is not under the control of the attacker.
To the attacker's benefit, though, the RSB is shared across processes running on the same logical CPU.
This allows the RSB to be poisoned from one process, and then be used by another process after a context switch.
For this attack to work, the attacker has to perform the following steps:
\begin{itemize}
	\item The attacker first fills the RSB with addresses of suitable code gadgets that leak secrets by creating a call instruction just before these gadgets' addresses and executing the call 16 times (step A2 from Section~\ref{sec:attack-general}).
	
    RSBs store virtual addresses of target instructions.
    Therefore, in order to inject the required address, we assume the attacker knows the target process's address space.
    In theory, in the case of a randomized address space (e.g., with ASLR), the attacker can use RSBs the opposite way, i.e., to leak the RSB entries, and thus to reveal the addresses of the victim process.
    However, we do not study this technique further in our evaluation.
	
	\item After filling the RSB, the attacker forces a context switch to the victim process (step A1 from Section~\ref{sec:attack-general}).
	For example, the attacker could call \code{sched\_yield} in order to ask the kernel to reschedule, ideally to the victim process.
    For this, we assume that the attacker process runs on the same logical CPU as the victim, and thus shares the RSB.
    This can be accomplished by changing the affinity of the process, to pin it to the victim's core (e.g., by using \code{taskset} in Linux), or alternatively, spawn one process per logical CPU.
\end{itemize}

\subsection{Proof-of-Concept Exploit}
To illustrate the general possibility of such cross-process data leaks, we picked a scenario where an attacker wants to leak user input, e.g., in order to capture user-entered passwords.
Thus, in our tested example, the victim is a terminal process that waits for user input, such as \code{bash} or \code{sudo}.
At a high level, such processes execute the following code:
\begin{ccode}
	while (inp = read_char(stdin)) {
		handle_user_input(inp);
	}
\end{ccode}

The following shows the (simplified) steps taken in a typical iteration of such an input-processing loop:
\begin{enumerate}
	\item The main loop starts.
	\item \code{read\_char} is called, which itself calls other intermediate functions, finally reaching the \code{read} system call.
	\item The \code{stdin} buffer will be empty (until the user starts typing) and the victim process is thus evicted.
	\item When the user presses a key, the victim process, waiting for buffer input, is scheduled.
	\item Execution continues within the \code{read} system call (which has just returned), and a chain of returns are executed until the execution reaches the main loop.
    \item \code{handle\_user\_input} handles the read character.
\end{enumerate}

In order to leak key presses, the attacker process has to be scheduled before the victim continues execution in (5).
This will guarantee that when resuming the execution from \code{read}, the victim process will use the attacker-tainted RSB entries.

\subsubsection{Leaking Key Presses}
\label{sec:attack-rsb-cross-process-keypress}

To show the plausibility of this attack, we implemented three programs:

\begin{description}
    \item[\texttt{Attacker}:] fills up RSB entries and preempts itself, so the victim process is scheduled after it.
    
    \item[\texttt{Measurer}:] probes for a predetermined set of memory pages to leak data from the victim process (using Flush+Reload~\cite{flush+reload}).
    
    \item[\texttt{Victim}:] simulates the behavior of \code{bash}, i.e., waits for new keystrokes and handles them when they are available.
    We replicate the call chain similarly to \code{bash} from the main loop to the \code{read} system call, and also return the same values.
\end{description}

There are several challenges with this approach:
\begin{enumerate}
    \item ASLR: Given that RSBs predict virtual addresses of jump targets, address space randomization in the victim makes tainting the RSB extremely difficult if not impossible.
    \item Small speculation window: Since we use memory accesses as a feedback mechanism, there is a race condition between adversarial memory access in speculation and reading the real return address off the stack.
    \item Post-Meltdown/Spectre (\cite{meltdown,spectre}) patches: RSBs have already been identified as a potential future risk that allows speculative execution. Thus, for CPU architectures that fall back to BTB upon RSB underflow, most modern OS kernels flush (technically, overwrite) RSBs every time a context switch occurs.
    \item Speculation gadgets: In \code{bash}, the character returned by the \code{read} system call is moved into \asm{rax} and the function returns.
    We targeted this return for speculation; thus, the required gadgets have to first shift \asm{rax} at the very least to a cache-line boundary (i.e., 6 bits to the left), and then do a memory access relative to shared memory (e.g., \asm{r10}, which contains the return address of the system call, pointing into libc: \asm{shl rax,6; mov rbx,[r10+rax]})
\end{enumerate}

Out of these challenges, (3) is a real limitation that cannot be avoided.
Flushing the RSB at context switches destroys the aforementioned attack.
To show that this prophylactic patch in modern OSes is indeed fundamental and needs to be extended to all affected CPUs, we for now assume RSBs are not flushed upon context switch.
Challenge (4) strongly depends on the compiler that was used to compile the victim program.
Therefore, each target requires its unique set of gadgets to be found.
Using an improved gadget finder, or by scanning various dynamic libraries, we believe the issue can be solved.
For our demo, we added the required gadgets.
Limitation (2) can be overcome by using another thread that evicts the addresses from the cache that correspond to the victim's stack.
In our experiments, for simplicity, we instead used \code{clflush} (an instruction invalidating the cache line of the provided address) to evict the victim's stack addresses in order to increase the speculation time window.
Finally, for limitation (1), we believe that our attack can be tweaked to derandomize ASLR of other processes.
For example, it could be possible to reverse the attack direction and cause the attacker process to speculate based on the victim's RSB entries to leak their value.
However, we do not evaluate this attack and assume that ASLR is either not deployed or there is an information leak that can be used to find the addresses of required gadgets in the victim process.

\subsection{Evaluation}
\label{sec:attack1-evaluation}

In the following, we evaluate the efficacy of our proof-of-concept implementation.
We carried out our experiments on Ubuntu 16.04 (kernel 4.13.0), running on Intel Haswell CPU (Intel\textsuperscript{\textregistered} Core\textsuperscript{\texttrademark} i5-4690 CPU @3.50GHz).
The Linux kernel, used in our evaluation, did not use RSB stuffing.

The execution environment was set according to the attack description in the previous section.
In the following, we note some implementation specifics.
So as to get rescheduled after each read key, our \texttt{Victim} process did not use standard input (\code{stdin}) buffering, which is in line with our envisioned target programs \code{bash} and \code{sudo}.
Additionally, a shared memory region is mapped in \texttt{Victim}, which will be shared with \texttt{Measurer}.
In our case, it was an \code{mmap}-ed file, but in reality this can be any shared library (e.g., \code{libc}).
In order to increase the speculation time, we used \code{clflush} in the \texttt{Victim}.
In practice, this has to be done by another thread that runs in parallel to \texttt{Victim} and evicts the corresponding cache lines of the \texttt{Victim}'s stack.
Finally, we also added the required gadget to \texttt{Victim}: \asm{shl rax,12; mov rbx,[r12+rax]}. 
At the point of speculative execution (i.e., when returning from \code{read}), \asm{rax} is the read character and \asm{r12} points to the shared memory region.

\texttt{Measurer} maps the shared (with \texttt{Victim}) memory in its address space, and constantly monitors the first 128 pages (each corresponding to an ASCII character).
In our experiments, we use Flush+Reload~\cite{flush+reload} as a feedback channel.
Finally, to be able to inject entries into \texttt{Victim}'s RSB, \texttt{Attacker} needs to run on the same logical CPU as \texttt{Victim}.
To this end, we modify both \texttt{Victim}'s and \texttt{Attacker}'s affinities to pin them to the same core (e.g., using \code{taskset} in Linux).
After that, \texttt{Attacker} runs in an infinite loop, pushing gadget addresses to the RSB and rescheduling itself (\code{sched\_yield}), hoping that \texttt{Victim} will be scheduled afterwards.

To measure the precision of our attack prototype, we determine the fraction of input bytes that \texttt{Measurer} read successfully.
To this end, we compute the Levenshtein distance~\cite{levenshtein-distance}, which measures the similarity between the source (S) and the destination (D) character sequences, by counting the number of insertions, deletions, and substitutions required to get D from S.
To measure the technique for each character in the alphabet, we used the famous pangram ``The quick brown fox jumps over the lazy dog''.
In the experiment, a new character from the pangram was provided to \texttt{Victim} every 50 milliseconds (i.e., 1200\,cpm, to cover even very fast typers).
Running the total of 1000 sentences resulted in an average Levenshtein distance of 7, i.e., an overall precision of $\approx$84\%.
It therefore requires just two password inputs to derive the complete password entered by a user using RSB-based speculative execution.

\section{Speculative Exec.~in Browsers}
\label{sec:attack2}

The cross-process attack presented in Section~\ref{sec:attack1} has demonstrated how a victim process might accidentally leak secrets via RSB-based misspeculations.
In this section, we consider a different setting with just a single process, in which a sandbox-contained attacker aims to read arbitrary memory of a browser process \emph{outside} of their allowed memory bounds.

\subsection{Threat Model}
\label{sec:attack2-threatmodel}
Scripting environments in web browsers have become ubiquitous.
The recent shift towards dynamic Web content has led to the fact that web sites include a plenitude of scripts (e.g., JavaScript, WebAssembly).
Browser vendors thus optimize script execution as much as possible.
For example, Just-in-Time (JIT) compilation of JavaScript code was a product of this demand, i.e., compiling JavaScript into native code at runtime.
Yet running possibly adversarial native code has its own security implications.
Multiple attacks have been proposed abusing JIT compilation for code injection~\cite{WCNBRCBL,dachshund,devil-in-constants,JITWX}.
Consequently, browser vendors strive to restrict their JIT environments as much as possible.
One such restriction, which our attack can evade, is sandboxing the generated JIT code such that it cannot read or write memory outside the permitted range.
For example, browsers compare the object being accessed and its corresponding bounds.
Any unsanitized memory access would escape such checks, and thus enable adversaries to read browser-based secrets or to leak data from the currently (or possibly all) open tabs, including their cross-origin frames.

In our threat model, we envision that the victim visits an attacker-controlled website.
The victim's browser supports JIT-compiled languages such as WebAssembly or JavaScript, as done by all major browsers nowadays.
We assume that the browser either has a high precision timer, or the attacker has an indirect timer source through which precise timing information can be extracted.

\subsection{WebAssembly-Based Speculation}
\label{sec:attack2-rsb-webassembly}

Our second attack scenario is also based on the general principles described in Section~\ref{sec:attack-general}.
However, in contrast to the scenario in Section~\ref{sec:attack1}, victim and target share the same process.
Furthermore, the attacker can now add (and therefore control) code that will be executed by the browser.
To this end, an attacker just has to ship JavaScript or WebAssembly code, both of which will be JIT-compiled by off-the-shelf browsers.
For illustration purposes and to have more control over the generated code, we focus on WebAssembly.

WebAssembly is a new assembly-like language, that is supported by all modern browsers (Edge from version 16, Chrome from 57, Firefox from 52, and Safari from 11.2)\footnote{https://caniuse.com/\#feat=wasm}.
As it is already low-level, compiling WebAssembly bytecode into native code is very fast.
The key benefit of WebAssembly is that arbitrary programs can be compiled into it, allowing them to run in browsers.
The currently proposed WebAssembly specification considers 4\,GiB accessible memory.
This makes sandboxing the generated code easier.
For example, in Firefox, usually a single register (\asm{r15} in x86) is dedicated as the pointer to the beginning of the memory, called the WebAssembly heap.
Consequently, all the memory is accessed relative to the heap.
To restrict all possible accesses into the 4\,GiB area, Firefox generates code that uses 32-bit x86 registers for encoding the offset.
Modifying a 32-bit register in x86-64 will zero the upper bits (e.g., \asm{add eax,1} will set the upper 32 bits of \asm{rax} to 0), thus limiting the maximum offset to 4~GiB.

\begin{figure}[t]
    \centering
    \includegraphics[bb=0 0 600 600,trim=0 60 220 15,clip,width=1.0\columnwidth]{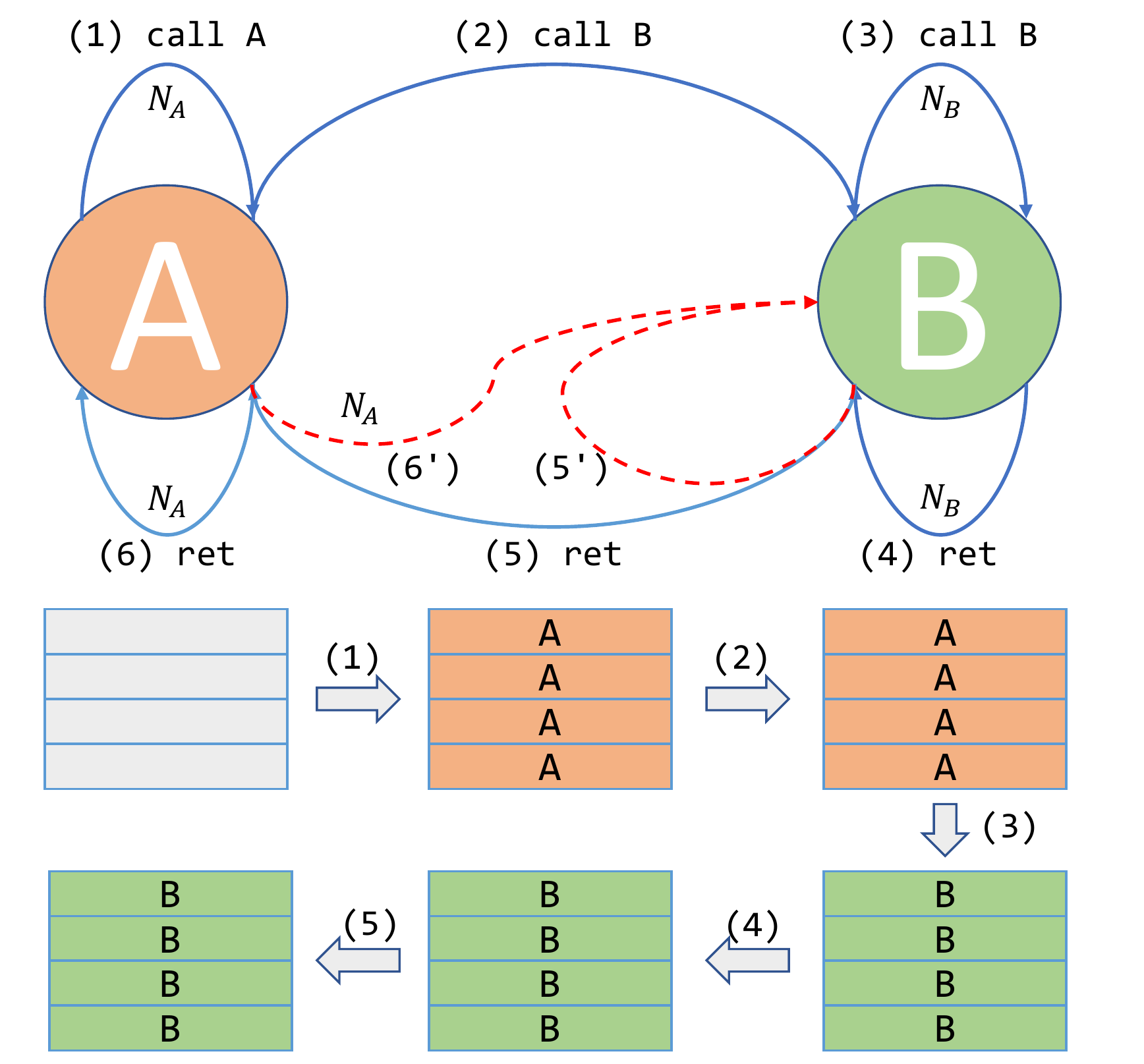}
    \caption{Cyclic RSB with recursive functions \texttt{A} and \texttt{B}. Dashed arrows show mispredicted returns, solid ones actual returns.}
    \label{fig:rsb-fsm}
\end{figure}

For our browser-based attack, we leverage cyclic RSBs to trigger misspeculation.
More precisely, we define two recursive functions $A$ and $B$, as shown in Figure~\ref{fig:rsb-fsm}.
In step (1), $A$ calls itself $N_A$ times and then calls $B$ in step (2).
In step (3), $B$ then calls itself recursively $N_B$ times, $N_B$ being the size of the RSB in this case.
The function $B$ follows two purposes.
First, being a recursive function, $B$ will overwrite all RSB entries with return addresses pointing to the instruction in $B$ following its recursive call.
Second, $B$ includes code right after calling itself to leak sensitive data using speculative execution in the context of $A$.
In step (4), $B$ returns $N_B$ times to itself, consuming all $N_B$ entries of the RSB.
However, since the RSB is cyclic, all the entries still remain there.
At this point, the return instruction in step (5) returns from $B$ to $A$ and triggers the first misprediction.
In step (6), $N_A$ more returns will be executed, all of them mispredicting $B$ as the return target.
The state of the RSB (shortened to $N=4$) after each of these steps is also depicted in Figure~\ref{fig:rsb-fsm}.

\subsection{Reading Arbitrary Memory}
\label{sec:attack2-rsb-webassembly-memread}

Compiling functions like those in Figure~\ref{fig:rsb-fsm} into WebAssembly bytecode will result in arbitrary speculation of the generated native code.
As a next step, we need to generate speculated code that leaks memory outside of the sandboxed memory region.
The key observation here is that whenever we trigger a misspeculation, we execute instructions of one function in the context of another.
For example, in the case of the functions \code{A} and \code{B} from Figure~\ref{fig:rsb-fsm}, after returning from \code{B} to \code{A}, code of function \code{B} will be speculatively executed, while the register values will stem from \code{A}.
This confusion of contexts allows evasion of defenses that are in place to sandbox JIT-compiled code.
As a simple example of context confusion, consider the following instruction accessing memory: \asm{mov rbx,[rax]}.
In normal execution, \asm{rax} will always be sanitized, e.g., by using 32-bit registers for offset calculation.
However, in speculation, triggered by another function (e.g., \asm{mov rax,0x12345678; ret}), \asm{rax} can be set to an arbitrary value, thus reading the data at an arbitrary memory location.

\begin{ccode}[label=lst:attack2-memory-read, caption=Arbitrary memory read in speculation, float, numbers=left]
uint8_t *B(int rec_N) {
    unsigned char *loc;
    if (rec_N > 0)
        loc = B(rec_N-1);
        // <-- speculation
    return &bytearray[bytearray[loc[0]<<12]];
}
uint64_t A(int rec_N) {
    uint_64 res = 0;
    if(rec_N > 0)
        res += A(rec_N-1);
        // <-- speculation context
    else
        res += *B(16);
    return ADDRESS; // attacker-controlled value
}
\end{ccode}

\begin{asmcode}[label=lst:attack2-memory-read-disasm, caption=Disassembly of functions A and B (important parts), float, numbers=left]
B:   ...
call B
mov  al, [r15+rax] ; r15=heap, rax=ADDRESS
shl  eax, 12       ; eax=leaked byte
mov  al, [r15+rax] ; report back the byte

A:   ...
mov  rax, ADDRESS
ret  ; trigger speculation in A, at line 3
; rax=ADDRESS will be used in speculation
\end{asmcode}

We will use these basic principles to generate speculative code that reads arbitrary memory contents---notably \emph{outside} of the sandboxed region.
To this end, we extend the general concept presented in Figure~\ref{fig:rsb-fsm} and derive WebAssembly code that emits the required instructions after compilation (Listing~\ref{lst:attack2-memory-read}).
The key concept here stays the same: function \code{A} calls itself recursively \code{rec\_N} times before calling \code{B}, which then recursively calls itself 16 times in order to fill up the RSB.
After 16 returns from \code{B}, \code{A} will return \code{rec\_N} times, each time triggering the speculation of instructions following the call statement in \code{B}, notably with the register contents of \code{A}.

The disassembly of the compiled functions \code{A} and \code{B} from Listing~\ref{lst:attack2-memory-read} are shown in Listing~\ref{lst:attack2-memory-read-disasm}.
After executing 16 returns from \code{B} (all with correct return address prediction), execution reaches the function \code{A}.
In \code{A}, the return value (\asm{rax}) is set (line 8) and the function returns (line 9).
At this point, as RSB was underflowed by executing 16 returns, the return address is mispredicted.
Namely, RSB's top entry will point to \code{B} (line 3).
While the correct return address is being read from the stack, lines 3 onwards are being executed speculatively.
The initial memory read operation (line 3) assumes a return value (\asm{rax}) to be set by \code{B}, which is supposed to be sanitized.
The base address, \asm{r15}, is a fixed pointer to WebAssembly's heap, which is also assumed to remain the same.
However, in our case, \asm{rax} was set in \code{A} with the attacker-controlled value.
This allows the attacker to read arbitrary memory relative to the WebAssembly heap.
Lines 4--5 are then used to report the value back by caching a value-dependent page.
That is, line 4 multiplies the read byte by 4096, aligning it to a memory page.
The page-aligned address is then used in line 5, where the $N$-th page of WebAssembly's heap is accessed.
After speculation, WebAssembly's heap can be probed to see which page was cached, revealing the leaked byte.

In our example, memory is leaked from an address relative to \asm{r15}, which points to WebAssembly's heap.
While the attacker-controlled offset (\asm{rax}) is a 64-bit register and covers the entire address space, it might still be desirable to read absolute addresses, e.g., in case one wants to leak the data from non-ASLRed sections of the memory.
This is easily doable with a simple modification of the WebAssembly bytecode.
Instead of using a direct call (\asm{call} opcode in WebAssembly), we can use an indirect call (\asm{indirect\_call} opcode).
The JIT compiler assumes that indirect calls might modify the \asm{r15} register, and therefore restores it from the stack when the callee returns.
Listing~\ref{lst:attack2-memory-read-disasm-ind} shows the disassembly of Listing~\ref{lst:attack2-memory-read-disasm} with this simple modification that added lines 3 and 4.
Line 3 restores the WebAssembly context register from the stack, while line 4 reads the heap pointer.
However, in speculative execution with \code{A}'s context, \asm{rsp} points to one of the arguments passed to \code{A}, which is controlled by the attacker.
Thus, the attacker controls the value of the heap pointer, and, by setting it to 0, can allow absolute memory accesses.

\begin{asmcode}[label=lst:attack2-memory-read-disasm-ind, caption=Disassembly of the function B with indirect call, float, numbers=left]
A:   ...
call rcx           ; rcx=A, dynamically set
mov  r14,[rsp]     ; rsp=&argN of B
mov  r15,[r14+24]  ; r15= argN of B
mov  al, [r15+rax] ; al = argN[ADDRESS]
shl  eax, 12       ; eax=leaked byte
mov  al, [r15+rax] ; report back the byte
\end{asmcode}

\subsection{Evaluation}
\label{sec:attack2-evaluation}

We now evaluate the efficacy and precision of our attack when applied for reading arbitrary memory in browsers.
We implemented our proof of concept in Firefox 59 on Windows 10 (version 10.0.16299), running on Intel's Haswell CPU (Intel\textsuperscript{\textregistered} Core\textsuperscript{\texttrademark} i5-4690 CPU @3.50GHz).
It is worth noting that Firefox, together with other browsers, has recently reduced the precision of performance counters to 2 milliseconds\footnote{https://developer.mozilla.org/docs/Web/API/Performance/now} as a defensive measure against Spectre~\cite{spectre}.
Given that finding alternative and more precise timing sources is out of the scope of this paper, we manually increased the performance counters to the old, more precise, state.

The main part of the proof of concept is a WebAssembly module that triggers the speculation.
The number of speculatively executed returns is customizable in the module by choosing a different recursion depth of function \texttt{A} ($N_A$); we set it to 64 return mispredictions in our experiments.
To feed back the speculatively read value, we used the WebAssembly heap of our module (from offset \code{0x4000} to avoid collision with other variables).
To avoid hardware prefetching interference, we access the heap at a page granularity, i.e., \code{Heap + 0x4000 + value*4096}.
After running the speculative code, we access the WebAssembly heap from JavaScript and measure the access times of each page.
Leaking the entire byte will require walking 256 memory pages, which would be very slow.
To optimize this, we split the byte in half (e.g., into \code{(value>>4)\&0xf} and \code{value\&0xf}) and leak each nibble separately.
This only requires scanning 16 pages per nibble, i.e., 32 scans per byte.
This could be further optimized to 8 per-bit reads.

Our measurements worked in the following order: 
(a) Using JavaScript, write the same pangram from Section~\ref{sec:attack1-evaluation} into a 1024-byte buffer.
(b) Compute the offset from the WebAssembly heap to the buffer containing the text.
(c) Trigger the eviction of the feedback cache lines from the cache, by doing random memory accesses to the same cache line in JavaScript.
(d) Call the WebAssembly module to speculatively execute the gadget from Listing~\ref{lst:attack2-memory-read-disasm}, reading the value from the specified offset.
(e) Scan the WebAssembly heap from JavaScript, and record the access times to each page.
(f) Repeat steps (c)--(e) 100 times to increase the confidence in the leaked data.
(g) Process the timings, recorded in (e), to find the page with the fastest average access time.
(h) Return the index of the found page.

In our evaluation, we ran each 1024-byte reading iteration 10 times.
Each iteration, on average, took 150\,seconds, i.e., $\approx$55\,bps reading speed---leaking a single byte thus takes 146\,ms.
Note that the main bottleneck in our measurements constitutes the code that evicts the cache lines (step (c)).
In our proof of concept, we simply map an L3 cache-sized buffer in JavaScript and then access each page to the corresponding cache line.
This approach can be further improved by initializing the eviction set prior to attack, and then walking the smaller set for eviction, as shown in~\cite{rowhammerjs}.

To measure the accuracy, similar to Section~\ref{sec:attack1-evaluation}, we used Levenshtein distance.
The evaluation showed that the read byte was correct $\approx$80\% of the time.
Increasing the iterations or number of speculations will increase the precision, however at the expense of reading speed.
We leave a more accurate and efficient implementation open for future work.
\section{Countermeasures}
\label{sec:discussion}

Seeing the immense impact of this new attack vector, in this section, we discuss countermeasures against RSB-based speculative execution.
Furthermore, we will describe the vendor reactions that followed our responsible disclosure process.

\subsection{Possible Mitigations}
\label{sec:discussion-mitigations}

In the following, we discuss possible mitigation techniques that can be employed to defend against our attacks.

\subsubsection{Hardware-based Mitigations}
\label{sec:discussion-hardware-defenses}

A naive approach to get rid of all speculative execution problems in hardware is to simply disable speculative execution.
That would, however, decrease performance drastically---making branch instructions serializing and forcing the execution of only a few instructions (between branches) at a time.
Of course, one could try to enable speculative execution while prohibiting speculative memory accesses, or at least caching them in speculation.
However, given that memory accesses are already a bottleneck for modern CPUs, blocking their speculative execution would incur a significant slowdown.

To counter our first attack in hardware, 
RSBs could be flushed by the CPU at every context switch.
Arguably, this will not impose any significant slowdown on performance, as the predictions after context switches will mispredict anyway in the vast majority of cases.
This is due to the fact that the RSB state is rarely shared between processes, as their virtual addresses are not the same (e.g., because of ASLR).
Furthermore, hardware-assisted flushing will be more efficient than a software-based solution that requires several artificially introduced calls (as implemented right now).
Hardware-backed RSB flushing would reliably prevent our cross-process attack, even in operating systems that do not flush RSBs themselves.

To counter our second attack, one could scrutinize the cyclic structure of RSBs and argue that switching to stack-based implementations mitigates the problem.
However, triggering a misspeculation is still possible in a size-bound (16-entry) RSB, e.g., by using exceptions in JavaScript, or relying on bailouts from JIT-compiled code (cf.~Section~\ref{sec:attack-general}).
We believe resorting to a combination of hardware/compiler solutions would allow more reliable security guarantees to defend against the second attack.

\subsubsection{Compiler-based Mitigations}
\label{sec:discussion-compiler-defenses}

To study how our second attack can be defended against in software, it is natural to ask how JIT compilers can be hardened.
Despite the fact that the general problem of speculative execution is caused by hardware, we can make our software environments more robust to these types of attacks.
The importance of this issue was recently highlighted, when multiple researchers proposed severe microarchitectural attacks, breaking down the core assumptions we had about hardware-based isolation and execution models~\cite{meltdown,spectre}.

For example, JIT compilers can aim to ensure that the code at call sites cannot be abused with any possible execution context.
The safest bet would be to stop all speculative executions at call sites, e.g., by using already-proposed solutions, such as \asm{lfence}/\asm{mfence} instructions (e.g., adding an \asm{lfence} instruction after every \asm{call} instruction).
Alternatively, one could introduce a modified version of a retpoline\footnote{https://support.google.com/faqs/answer/7625886} that replaces all return instructions emitted by JIT compilers by a construct that destroys the RSB entry before returning:

\begin{asmcode}[numbers=left]
  call return_new ;
speculate:        ; this will speculate
  pause           ; trap speculation until...
  jmp  speculate  ; ...return address is read
return_new:       ;
  add  rsp, 8     ; return to original addr.
  ret             ; predict to <speculate>
\end{asmcode}
\noindent
The call instruction (line~1) guarantees that the RSB will have at least one entry and will not underflow.
Lines~6-7 then make sure that the architectural control flow will continue from the original return address (\asm{rsp+8}), while the speculative one will be trapped in the infinite loop (lines~2-4).

Alternatively, one could improve the memory access sanitization in JIT compilers.
For example, JIT-compiled code could always use 32-bit registers as a natural way to constrain addresses to a 4\,GiB range in memory---the current memory limit in WebAssembly.
However, this by itself does not provide strong security guarantees.
As we have shown in Section~\ref{sec:attack2-rsb-webassembly-memread}, the base addresses can also be modified in speculation.
Having said this, WebAssembly is a relatively new addition to browsers, and new features are still being frequently suggested/developed.
Each of these feature needs to be reevaluated in our context.
In particular, the proposals to add exception handling and threading support to WebAssembly need to be carefully revisited.
Built-in exception handling will allow RSB speculation even with a non-cyclic RSB, while adding WebAssembly support for threading might introduce new precise timing side channels.

Regardless of the precise countermeasure, one can limit the overhead of compiler-based defenses.
In particular, code parts that are guaranteed to be secure against all potential abuses (e.g., possibly speculated code that does not have memory accesses) can be left as is.

\subsubsection{Browser-based Mitigations}
\label{sec:discussion-timing-defenses}

One of the directions that browser vendors take to mitigate side-channel attacks is to deprive the attackers of precise timings.
Having no timers, adversaries cannot distinguish between cached and non-cached memory accesses, which is a fundamental requirement for cache- and timing-based side-channel attacks.
Given the complexity of JavaScript environments, merely decreasing the \code{performance.now} counter (as done in most browsers) is insufficient.
For example, \etal{Gras}~\cite{aslr_break_AnC} showed that \code{SharedArrayBuffer} can be used to acquire a timing source of nanosecond precision, while \etal{Schwarz}~\cite{fantastic_timers} studied different timing sources in modern browsers, ranging from nanosecond to microsecond precision.
Approaches presented in academia thus aim to advance this protection to the next level.
For example, the ``Deterministic Browser'' from \etal{Cao}~\cite{deterministic_browser} tries to tackle the issue by proposing deterministic timers, so that any two measurements from the same place will always result in the same timing value, thus making it useless for the attacker.
In another study, Kohlbrenner and Shacham~\cite{fuzzyfox} propose Fuzzyfox, which aims to eliminate timers by introducing randomness, while also randomizing the execution to remove indirect time sources.
Motivated by these works, and by the recent discovery of Spectre, browsers decreased their timing precision to 2 milliseconds, while also introducing a jitter, such that the edge thresholding technique shown by \etal{Schwarz}~\cite{fantastic_timers} is also mitigated.

Alternatively, browsers can alleviate the threats by stronger isolation concepts.
In the most extreme case, browsers can add dedicated processes for each entity (e.g., per site) to enforce a strict memory separation and to isolate pages from each other.
By doing so, one can guarantee that even with a severe vulnerability at hand, such as arbitrary memory read, adversaries are constrained to read memory of the current per-page process.
Reportedly, modern browsers already consider this technique, e.g., Chrome uses a dedicated sandboxed process per domain~\cite{chrome-siteisolation-document}, while Firefox plans to switch to a similar architecture in the near future.

\subsection{Responsible Disclosure}
\label{sec:disclosure}

Seeing the severity of our findings, we have reported the documented attacks to the major CPU vendors (Intel, AMD, ARM), OS vendors (Microsoft, Redhat) and browser developers (Mozilla, Google, Apple, Microsoft) in April 2018, and subsequently engaged in follow-up discussions.
In the following, we will summarize their reactions and our risk analysis.

\textbf{Intel:}
Intel acknowledged this ``very interesting'' issue of RSB-based speculative execution and will further review the attack and its implications.
Their immediate advice is to resort to mitigations similar to Spectre is to defend against our attack (see Section~\ref{sec:discussion-mitigations}); this is, however, subject to change as part of their ongoing RSB investigations that we triggered.

\textbf{Mozilla Foundation:}
The Mozilla Foundation likewise acknowledged the issue.
They decided to refrain from using compiler-assisted defenses, as they would seemingly require complex changes to JIT-compiled and C++ code.
Instead, they aim to remove all (fine-granular) timers from Firefox to destroy caching-based feedback channels.
Furthermore, they referred to an upcoming Firefox release that includes time jittering features similar to those described in FuzzyFox~\cite{fuzzyfox}, which further harden against accurate timers.

\textbf{Google:}
Google acknowledged the problem in principle also affects Chrome.
Similar to Firefox, they do not aim to address the problem with compiler-assisted solutions.
Instead, they also refer to inaccurate timers, but more importantly, focus on a stronger isolation between sites of different origins.
Chrome's so-called Site Isolation prevents attackers from reading across origins (e.g., sites of other domains).

\textbf{AMD / ARM:} Although we have not tested our attacks against ARM and AMD architectures, they acknowledged the general problem.

\textbf{Microsoft:} Microsoft has acknowledged the problem and is working on fixes, but has not disclosed technical details yet.

\textbf{Apple:} As of 08/15/2018, we have not heard back from Apple yet.

\textbf{Redhat:} Redhat was thankful for our disclosure and said that the current Spectre defenses such as flushing RSBs---without considering RSB-based attacks---might otherwise have been removed by the kernel developers in the near future. In particular, Redhat stressed that fixing RSB underflows will not fully solve the problems we point out in our paper.

\section{Related Work}
\label{sec:relatedwork}

In the following, we discuss concepts related to our paper.
First, we provide an overview of the two recent papers on speculative and out-of-order executions that are both closest to our work.
We will then briefly summarize other similar work done in that area.
Further, we look into microarchitectural attacks in general, discussing some notable examples.
Finally, we also discuss proposed defense techniques and their efficacy against our proposed attacks.

\subsection{Out-of-Order/Speculative Execution}
Despite being implemented in CPUs since the early 90s, out-of-order and speculative executions have only recently caught the attention of security researchers.
Fogh was the first to document speculative execution and reported his (negative) findings~\cite{anders-fogh-negative-results}.
Spurred by this general concept, multiple researchers then ended up discovering similar bugs~\cite{meltdown,spectre,speculose} more or less simultaneously.
In the following, we summarize the principles behind two representative candidates from this general attack class, namely Meltdown and Spectre.

On the one hand, Meltdown~\cite{meltdown} (a.k.a.~Variant 3) abuses a flaw in Intel's out-of-order execution engine, allowing adversaries to have access to data for a split-second without checking the privileges.
This race condition in the execution core allows attackers to disclose arbitrary privileged data from kernel space.
While it is the most severe, Meltdown is relatively easy to counter with stronger separation between user and kernel space~\cite{kaiser}.

Spectre~\cite{spectre} (a.k.a.~Variants 1 and 2), on the other hand, does not rely on implementation bugs in CPUs, and is therefore also significantly harder to tackle.
Technically, Spectre uses a benign CPU feature: speculative execution.
The problem, however, is that the branch predictor is shared between different processes and even between different privileges running on the same core.
Therefore, in Spectre, adversaries are able to inject arbitrary branch targets into predictors and, by doing so, trigger arbitrary speculative code execution in victim processes (similar to our first attack).
Furthermore, similar to our second attack, Spectre also proposed an in-browser attack to abuse the branch predictor in the same process, where mispredicting a branch path can lead to leakage of unauthorized data.
Spectre is thus closely related to our approach.
The difference is that we achieve similar attack goals by abusing a completely different prediction mechanism of the CPU: return stack buffers.
While RSBs were already mentioned as a \emph{potential} security risk~\cite{meltdown-p0,spectre}, it was so far unclear whether RSBs indeed pose a threat similarly severe as BTBs.
Our work answers this open question and provides countermeasures to this problem.

Follow-up works naturally arose out of the general discovery of Meltdown and Spectre.
In SgxPectre, for example, \etal{Chen}~\cite{sgxpectre} showed that it is possible to use branch target injection to extract critical information from an SGX enclave.
Similarly, in BranchScope, \etal{Evtyushkin}~\cite{branchscope} studied the possible abuses of direct branch predictors to leak sensitive data from different processes, including SGX enclaves.

\subsection{Cache-Based Side Channels}
\label{sec:relatedwork-cache-sidechannels}

Given that accessing main memory in modern CPUs can take hundreds of cycles, current architectures employ multiple layers of caches.
Each layer has various characteristics and timing properties, thus providing unique information as side channels.
The key idea of cache side channel attacks is to distinguish the access times between cache hits and misses, revealing whether the corresponding data was cached or not.

Cache attacks can be divided into attacks on instruction and data caches.
The attacks on instruction caches aim to leak information about the execution of the target program.
For example, information from instruction caches can be used to reconstruct the execution trace of colocated programs~\cite{yet-another-microarch-attack-icache,new-results-icache,rsa-vuln-icache,improve-rsa-attack-icache} or even VMs on the same machine~\cite{cross-vm-icache}.

In contrast, side channels from data caches reveal data access patterns in the target program, which again can be either a colocated program or a VM, depending on the level of the attacked cache.
Per-core caches (e.g., L1 and L2) can be used as side channels against programs running on the same physical core.
This has been shown to be useful for reconstructing cryptographic keys~\cite{efficient-cache-attacks-dcache}.
Conversely, shared last-level caches (LLC) can be used to leak information, e.g., keystrokes or user-typed passwords, from any process running on the same CPU---notably even across VMs~\cite{hey-get-off-my-cloud}.

There are different ways to leak data via caches.
Most notably, Flush+Reload~\cite{flush+reload} uses \asm{clflush} to flush the required cache lines from all cache levels including the last-level cache shared with the victim.
By measuring the access time to the same cache line, the attacker can detect whether the victim has accessed a certain cache line.
Some variations of the Flush+Reload attack include Evict+Reload~\cite{cache-template-attacks}, which tries to evict the target cache line by doing memory accesses instead of the \asm{clflush} instruction.
This is important for cases where \asm{clflush} cannot be used, e.g., in JIT code (cf.~Section~\ref{sec:attack2}), or architectures without an instruction similar to \asm{clflush}~\cite{armageddon}.
The inverse of Flush+Reload is Prime+Probe~\cite{prime+probe}, where the adversary allocates (primes) the entire cache (or a specific cache set) with its own data, and then triggers the execution of the victim.
Then, the attacker will probe the cache lines to see which of them have been evicted (i.e., which cache lines have been accessed) by the victim.

\subsection{Other Microarchitectural Side Channels}
\label{sec:relatedwork-cpu-sidechannels}

Given the complexity and abundance of optimizations, side channels in microarchitectures is not surprising anymore.
Therefore, there are plenty of different attack techniques proposed by researchers, each of which target microarchitectural features of modern CPUs.
For example, \etal{Evtyushkin}~\cite{kaslr_break_branchpredictor} use collisions in BTBs to leak information about the kernel address space, and by doing so derandomize the kernel-level ASLR (KASLR).
Similar to Meltdown, which uses out-of-order execution to suppress exceptions, \etal{Jang}~\cite{kaslr_break_tsx} use Intel's transactional synchronization extensions (TSX).
By accessing kernel pages with TSX, depending on the type of the generated exception (e.g., a segmentation fault if memory is unmapped, or a general protection fault if the process does not have the privileges to access certain memory regions), the time to roll back the transaction differs.
This constitutes a timing side channel that can be used to bypass KASLR, as an attacker can probe pages mapped in the kernel.
Similarly, \etal{Gruss}~\cite{kaslr_break_prefetch} measure timing differences between prefetching various kernel-land memory addresses to distinguish between mapped and unmapped addresses.
Finally, \etal{Hund}~\cite{kaslr_break_practicaltiming} propose three different attack techniques to derandomize KASLR: Cache Probing, Double Page Fault, and Cache Preloading.

\subsection{ret2spec vs.~Spectre Returns}
The general attack idea of our paper was also discovered by \etal{Koruyeh}~\cite{spectre-rsb}, who called the attack \emph{Spectre Returns}.
The authors of \emph{Spectre Returns} describe the general problem with RSBs and propose a few attack variations that are similar to ours described in Section~\ref{sec:attack1}.
Additionally, they proposed using RSB poisoning to trigger speculative execution in SGX.
Instead, we target JIT environments to read arbitrary memory in modern browsers, an attack that even works in presence of RSB flushes.
We would like to highlight that our paper submission (to ACM CCS '18) predates the one of \emph{Specture Returns} (to both USENIX WOOT '18 and Arxiv), as also illustrated by the fact that our responsible disclosure started already in April 2018.
\section{Conclusion}
\label{sec:conclusion}
In this work, we investigate the security implications of speculative execution caused by return stack buffers (RSBs), presenting general principles of RSB-based speculative execution.
We show that RSBs are a powerful tool in the hands of an adversary, fueled by the simplicity of triggering speculative execution via RSBs.
We demonstrate that return address speculation can lead to arbitrary speculative code execution across processes (unless RSBs are flushed upon context switches).
Furthermore, we show that in-process speculative code execution can be achieved in a sandboxed process, resulting in arbitrary memory disclosure.

\section*{Acknowledgments}
The authors would like to thank the anonymous reviewers for their valuable comments.
Moreover, we would like to thank Fabian Schwarz for his feedback and support during the writing process of the paper.

This work was supported by the German Federal Ministry of Education and Research (BMBF) through funding for the CAMRICS project (FKZ 16KIS0656).

\bibliographystyle{ACM-Reference-Format}
\bibliography{paper}

\end{document}